\documentclass[11pt,a4paper]{article}

\usepackage[T1]{fontenc}
\usepackage[utf8]{inputenc}

\usepackage{amsmath,amssymb,amsthm,bm}
\usepackage{mathtools}

\usepackage{mathptmx}     % Times-like serif (swap for lmodern if preferred)
\usepackage{microtype}

\usepackage[a4paper,margin=1in]{geometry}
\usepackage{setspace}
\usepackage{titlesec}
\usepackage{enumitem}
\usepackage{graphicx}
\usepackage[font=small,labelfont=bf,margin=1em]{caption}

\usepackage{xcolor}
\usepackage[colorlinks=true,
            linkcolor=black,
            citecolor=blue!55!black,
            urlcolor=blue!60!black,
            bookmarksopen=true,
            pdftitle={The Saddle Point of Everything},
            pdfauthor={K. Sravan Kumar}]{hyperref}
\usepackage{cite}

\titleformat{\section}
  {\normalfont\large\bfseries\raggedright}{\thesection.}{0.6em}{}
\titleformat{\subsection}
  {\normalfont\normalsize\bfseries\itshape\raggedright}{\thesubsection.}{0.5em}{}
\titlespacing*{\section}{0pt}{1.6\baselineskip}{0.6\baselineskip}
\titlespacing*{\subsection}{0pt}{1.0\baselineskip}{0.3\baselineskip}

\newcommand{\MP}{M_{\mathrm{P}}}
\newcommand{\Wmnrs}{W_{\mu\nu\rho\sigma}}
\newcommand{\sqmg}{\sqrt{-g}}
\newcommand{\PV}{\mathrm{PV}}
\newcommand{\IHO}{\mathrm{IHO}}
\newcommand{\dIHO}{\mathrm{dual\text{-}IHO}}

\newtheoremstyle{thm-saddle}%
  {1.0\baselineskip}% above
  {0.4\baselineskip}% below
  {\itshape}% body font
  {}% indent
  {\bfseries}% header font
  {.}% punctuation after head
  {0.5em}% space after head
  {}% custom head spec
\theoremstyle{thm-saddle}
\newtheorem{theorem}{Theorem}

\begin{document}

% ============================================================
%  Title block
% ============================================================
\begin{center}
  {\LARGE\bfseries The Saddle Point of Everything}\\[10pt]
  {\large\itshape The inverted harmonic oscillator,\\
   the universal physics of unstable equilibria,\\
   and the unique renormalizable theory of quantum gravity}\\[18pt]
  {\large K.\ Sravan Kumar}\\[4pt]
  {\small\itshape Institute of Cosmology and Gravitation,
   University of Portsmouth,\\
   Dennis Sciama Building, Burnaby Road, Portsmouth PO1\,3FX, United Kingdom}\\[2pt]
  {\small\ttfamily sravan.kumar@port.ac.uk}\\[10pt]

\end{center}

\bigskip

% ============================================================
%  Abstract  --  preserved verbatim, do not edit.
% ============================================================
\noindent\textbf{Abstract.} The harmonic oscillator is the universal Hamiltonian of stable equilibrium. Its counterpart, the inverted harmonic oscillator (IHO), is the Hamiltonian of unstable equilibrium: the saddle point of physical systems. It appears across disciplines, from condensed matter, quantum optics, and quantum chemistry to the Standard Model Higgs instability and quantum field theory near gravitational horizons. Its mathematical depth is further reflected in its relation to the non-trivial zeros of the Riemann zeta function through the Berry--Keating Hamiltonian. Remarkably, a dual Hamiltonian to the IHO has recently been shown to govern the additional spin-2 sector of the unique unitary perturbatively renormalizable theory of quantum gravity in four dimensions, with that sector remaining purely virtual and regularizing gravitational interactions at the Planck scale. This paper argues that the universal physics of the saddle point course-corrects the history of quantum gravity approaches that abandoned renormalizability, uniqueness, and predictivity. Its consequences include a non-singular Universe, Starobinsky inflation, and possible implications for large-scale CMB features and primordial gravitational waves.

\medskip


\bigskip
\hrule
\bigskip

% ====================================================
\section{The harmonic oscillator built physics. What did its inverted counterpart build?}
\label{sec:opening}

The first equation a physicist learns to take seriously is the harmonic oscillator. Expand any potential $V(q)$ around a minimum, retain the leading non-trivial term, and one obtains
\begin{equation}
\label{eq:HO}
H \;=\; \frac{p^{2}}{2m} \;+\; \frac{1}{2}m\omega^{2}q^{2}.
\end{equation}
This is not a special example. It is universal. Every theory of stable equilibrium, such as molecular vibration, lattice phonons, blackbody radiation, the Dirac vacuum of quantum electrodynamics (QED), the perturbative vacuum of the Standard Model, is built, at leading order, on harmonic oscillators stacked together. The Fock-space formalism of perturbative quantum field theory is, in essence the systematic exploitation of this fact: take a Lorentz-invariant potential energy that has a minimum, quantize the small fluctuations as a sum of independent oscillators, and label the excitations as particles. The harmonic oscillator is so central to physics that the question of what \emph{other} elementary system might play a comparable role rarely gets asked.

The natural counterpart is the inverted harmonic oscillator (IHO). Expand the same $V(q)$ around a maximum, retain the leading non-trivial term, and one obtains
\begin{equation}
\label{eq:IHO}
H_{\IHO} \;=\; \frac{p^{2}}{2m} \;-\; \frac{1}{2}m\omega^{2}q^{2} \;=\; \frac{\omega}{2}\bigl(\tilde p^{\,2}-\tilde q^{\,2}\bigr).
\end{equation}
The sign of the second term reverses. The consequences are dramatic. The Hamiltonian is now indefinite, neither bounded above nor below. The eigenfunctions are parabolic-cylinder functions $D_{\nu}(\tilde q)$ that grow exponentially in one asymptotic direction and decay in the other. The spectrum is real and continuous; eigenstates are Dirac-delta normalizable but not square integrable. There is no normalizable ground state. Under the rotation $Q=(\tilde p+\tilde q)/\sqrt{2}$, $P=(\tilde p-\tilde q)/\sqrt{2}$, the same Hamiltonian assumes its Berry--Keating form,
\begin{equation}
\label{eq:HBK}
H_{\IHO} \;=\; \frac{\omega}{2}\,(QP+PQ),
\end{equation}
the generator of dilatations. Classical trajectories are hyperbolae, $Q(t)=Q_{0}e^{\omega t}$, $P(t)=P_{0}e^{-\omega t}$. The phase space is divided by separatrices ($\tilde p=\pm \tilde q$) into four regions and the flow within each is asymptotic to a separatrix. The mathematical and physical properties of this elementary object were systematized by Barton in 1986 \cite{Barton1986}, and have since been re-derived, often without cross-referencing, at every saddle point and horizon of modern theoretical physics.

\begin{figure}[t]
  \centering
  \includegraphics[width=0.95\linewidth]{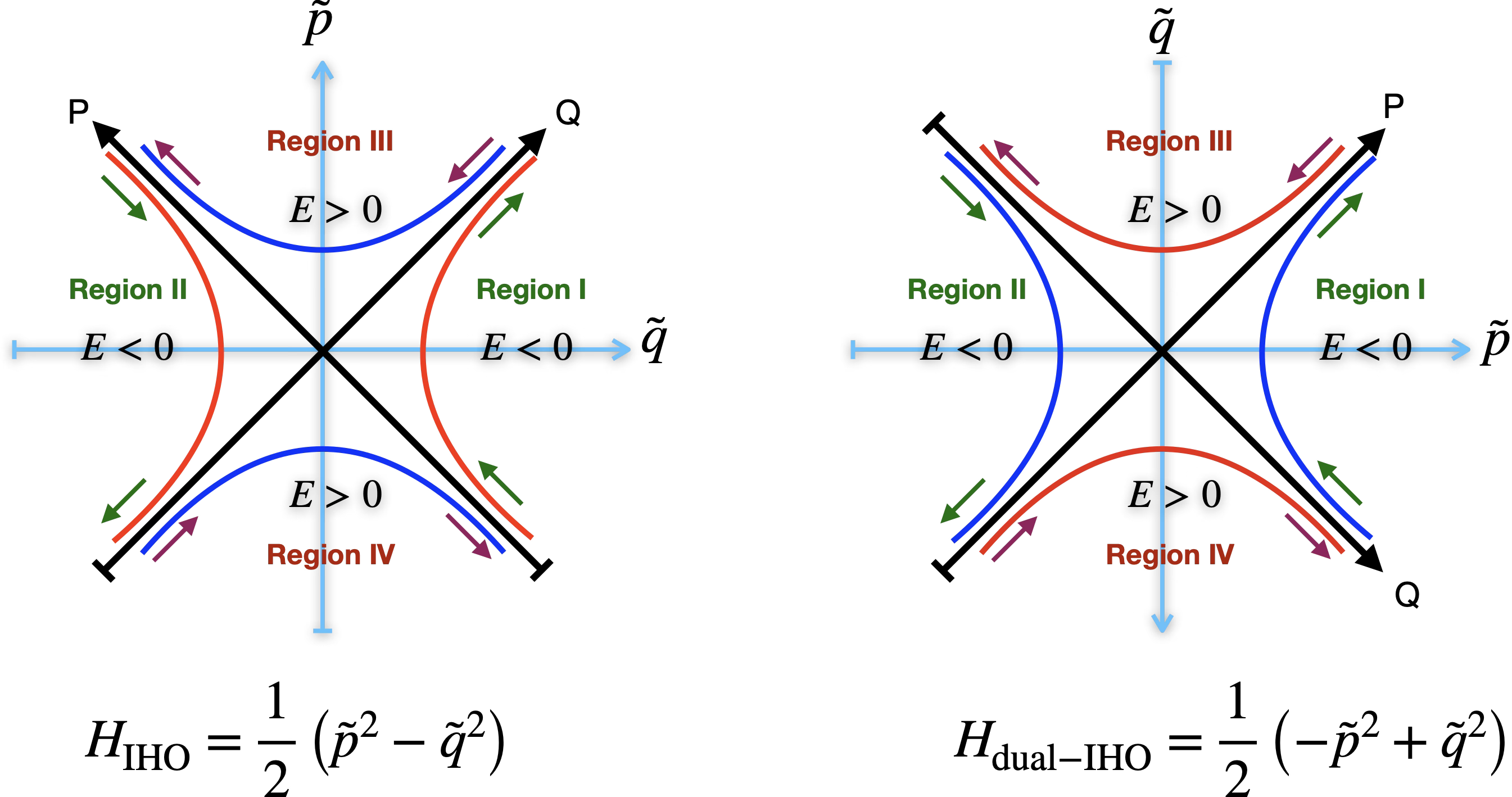}
  \caption{Phase-space portrait of the inverted harmonic oscillator (left) and
    its dual partner (right). In each panel the hyperbolic separatrices
    divide the $(Q,P)$ plane into four regions (I--IV); classical
    trajectories are asymptotic to the separatrices and the flow in each
    region carries opposite-sign Lyapunov exponents along the two principal
    directions. The dual-IHO is obtained from the IHO by sign reversal of
    both the kinetic and the potential terms in
    Eq.~\eqref{eq:Hslambda}; at the level of the classical flow, this is a
    time reversal acting on Regions~III and IV. The four-region structure
    is the geometric origin of the direct-sum Hilbert space
    $\mathcal H_{+}\oplus\mathcal H_{-}$ on which the unitary
    quantization of the dual-IHO sector of quadratic gravity is built
    (Section~\ref{sec:KL}).}
  \label{fig:iho-dual}
\end{figure}

An equivalent physical system to the IHO is its dual,
\begin{equation}
\label{eq:dIHO}
H_{\dIHO} \;=\; \frac{\omega}{2}\bigl(-\tilde p^{\,2}+\tilde q^{\,2}\bigr),
\end{equation}
which is again an indefinite hyperbolic Hamiltonian, obtained from the IHO by simultaneous sign reversal of the kinetic and potential terms or by the transformation $\tilde p\leftrightarrow \tilde q$. The IHO and the dual-IHO share equivalent phase spaces related by $90^\circ$ rotation. They together exhaust the four-region saddle structure of Figure~\ref{fig:iho-dual}. This dual form will be the relevant one when we turn, in Section~\ref{sec:stelle}, to the massive spin--2 sector of quadratic gravity.

% ---- Figure: IHO / dual-IHO phase portrait ----
The IHO field should not just be seen as a "tachyon", which, in the conventional flat-space sense, is often considered to be a putative particle of imaginary mass with an asymptotic state and definable (unstable) vacuum \cite{Jacobson:1987ap}. 
The dual-IHO field is also not a ghost. A ghost that is associated with a negative definite Hamiltonian is a Hilbert-space pathology: a state whose norm is negative under the inner product, leading directly to violations of the Born rule. Both IHO and dual-IHO fields can be defined with positive norms, their Hamiltonian is real and indefinite, and inner products can be the standard ones. What they lack is a normalizable vacuum, and what their eigenstates lack is an ordinary Fock-particle interpretation. This is a different thing from pathology. The next section will show that it is the relevant description of an extraordinary range of physics.

This paper makes two connected claims. First, that the IHO is the universal Hamiltonian of \emph{instability} in the same precise sense that the harmonic oscillator is the universal Hamiltonian of stability. It is what physics encounters at every saddle point, every horizon, every transition state, every critical point, every direction of genuine instability across condensed matter, atomic and molecular physics, gravity, cosmology, and mathematical physics. Second, that the dual IHO is the correct identification of the additional spin--2 sector of the unique perturbatively renormalizable theory of gravity in four dimensions. The decades-long verdict that this theory was non-unitary rested on identifying that sector as a massive ghost that leads to a particle excitation of a negative-definite Hamiltonian rather than as a dual IHO, an indefinite-Hamiltonian instability with no particle interpretation at all. Once the distinction is admitted, renormalizability and unitarity coexist in the original local four-dimensional action that Stelle wrote down in 1977 \cite{KumarMartoQQG2026, KumarMartoUnitarity2026}.

These two claims are not independent. They are the same claim, separated only by the systems they are applied to. To find the IHO at the heart of quantum gravity is to find that quantum gravity speaks the same elementary saddle-point language that the rest of physics already speaks.

% ====================================================
\section{The inverted harmonic oscillator as a universal saddle structure}
\label{sec:universality}

This section traces the IHO through a sequence of physical systems. The claim is empirical, not metaphysical: the IHO is the locally accurate Hamiltonian of a saddle point, and saddle points are everywhere. Where the harmonic oscillator describes the elementary physics of \emph{return} (small fluctuations dissipate or oscillate back), the IHO describes the elementary physics of \emph{transition} (small fluctuations grow until the system reorganizes around a new structure).

\subsection{The Mexican hat: IHO at the symmetric point of the Standard Model}

The Higgs sector is the most familiar place where the IHO sits at the heart of established physics \cite{EnglertBrout1964, Higgs1964}. The potential
\begin{equation}
\label{eq:Higgs}
V_{H}(H) \;=\; -\,\frac{\mu_{H}^{2}}{2}\,H^{\dagger}H \;+\; \frac{\lambda_{H}}{4}\bigl(H^{\dagger}H\bigr)^{2}
\end{equation}
has, at the origin, a tachyonic mass term: a negative coefficient that makes the symmetric point an unstable maximum. The local Hamiltonian at the origin is, in each transverse field direction, an inverted oscillator. The phase transition that breaks electroweak symmetry is driven by this IHO: small fluctuations of $H$ at the symmetric point grow exponentially, the field rolls outward to $|H|=v=\sqrt{\mu_{H}^{2}/\lambda_{H}}$, and only there, expanding around the true vacuum, does ordinary harmonic-oscillator physics resume in the form of the physical Higgs boson and the longitudinal modes of the $W^{\pm}$ and $Z^{0}$.

The same structure governs every Landau--Ginzburg phase transition near its critical temperature \cite{LandauGinzburg1950, Hohenberg1977}. Above $T_{c}$ the order parameter sits at a stable equilibrium; below $T_{c}$ the symmetric point becomes an IHO and the system undergoes a coarsening transition into broken-symmetry domains. Critical slowing down, the divergence of the correlation length, the universality classes that organize second-order transitions---all are governed by IHO dynamics at criticality.

The Standard Model is, in this sense, an IHO-driven theory in its very foundation. The Higgs mechanism is a triumph of the inverted harmonic oscillator that gets recategorized, after the fact, as a triumph of the harmonic one because the post-symmetry-breaking physics is what we observe. The structure that drives the transition is silent in the spectrum it produces, and so is mostly forgotten.

\subsection{Transition states: the IHO of every chemical reaction}

In quantum chemistry, the rate of a reaction is governed by the saddle point of the potential energy surface that separates reactants from products. Wigner derived the first quantum tunneling correction at such a saddle in 1932 \cite{Wigner1932rate}. The local Hamiltonian at the transition state, in the reaction-coordinate direction $q$, is precisely an inverted oscillator,
\begin{equation}
\label{eq:Hrxn}
H_{\mathrm{rxn}} \;=\; \frac{p^{2}}{2\mu} \;-\; \frac{1}{2}\kappa^{\ddagger}\,q^{2},
\end{equation}
with imaginary barrier frequency $\omega^{\ddagger}=\sqrt{\kappa^{\ddagger}/\mu}$. Miller's semiclassical transition-state theory \cite{Miller1974}, Grote--Hynes theory for non-equilibrium solvation \cite{GroteHynes1980}, and the Pollak--Grabert--H\"anggi formulation of dissipative barrier crossing \cite{PollakGrabertHanggi1989} all reduce, at their analytic core, to the IHO propagator at the dividing surface. Every numerical computation of a chemical rate in a regime where tunneling matters uses, somewhere in its derivation, the parabolic-cylinder Green's function of the inverted oscillator. The Eyring equation that physical chemists learn as undergraduates \cite{Eyring1935} is, in a real sense, the leading classical limit of an IHO calculation.

\subsection{Condensed matter: the lowest Landau level and the unification across communities}

Subramanyan, Hegde, Vishveshwara, and Bradlyn established in 2020--21 \cite{Subramanyan2020} that the scattering of lowest-Landau-level (LLL) electrons through a quantum point contact in a perpendicular magnetic field is governed exactly by an inverted harmonic oscillator. A two-dimensional electron gas in a strong field organizes into Landau levels separated by $\hbar\omega_{c}$, and within the LLL, the scattering through a saddle-point contact which is the prototype of integer quantum Hall edge physics is governed by
\begin{equation}
H_{\mathrm{LLL}} \;=\; \frac{1}{2}m^{\ast}\omega_{c}^{2}\,(x^{2}-y^{2}),
\end{equation}
an IHO in the area-preserving (dilatation) form of \eqref{eq:HBK}. The transmission coefficient through the contact takes the form
\begin{equation}
\label{eq:TLLL}
T(E) \;=\; \frac{1}{1+e^{-2\pi E/\hbar\omega_{c}}},
\end{equation}
the Fermi--Dirac distribution centered at the classical turning energy. This is exactly the same spectrum that emerges from the Bogoliubov decomposition of Minkowski vacuum modes seen by a Rindler observer (the Unruh effect) and from the Hawking calculation of black-hole evaporation. The transmission spectrum of a quantum point contact in the lowest Landau level and the Hawking temperature of a black hole are, mathematically, the same calculation: the thermal IHO scattering matrix in its area-preserving form, with $\omega_{c}$ in one case and the surface gravity $\kappa$ in the other.

The Subramanyan et al.\ paper is, in this respect, one of the cleanest existing demonstrations that the IHO unifies fields of physics usually thought of as belonging to separate communities. It is among the strongest pieces of evidence that the IHO is not a quirky toy model used in special technical contexts, but a universal physical structure. The argument of the present paper is, in part, that this universality extends one further step, that is, into the ultraviolet sector of quantum gravity itself.

\subsection{Schwinger pair production and the Sauter potential}

In a constant external electric field $\mathcal{E}$, virtual electron--positron pairs experience an effective barrier in momentum space whose shape, after appropriate canonical transformation, is an inverted parabola. The Schwinger pair production rate
\begin{equation}
\Gamma \;\propto\; \exp\!\Bigl[-\,\frac{\pi\,m_{e}^{2}}{e\mathcal{E}}\Bigr]
\end{equation}
is computed by IHO barrier penetration \cite{Schwinger1951}. The exact Sauter pulse \cite{Sauter1932}, with field profile $\mathcal{E}(t) = \mathcal{E}_0\,\mathrm{sech}^2(t/\tau)$, is a smooth finite-width version of the constant field. It admits an analytically solvable IHO scattering problem whose reflection and transmission coefficients reduce to the Schwinger result in the constant-field limit $\tau\to\infty$. Recent atomic and molecular physics proposals to observe analog Schwinger pair
production in cold atomic systems \cite{ZacheEtAl2019}, and the active research
program on dynamical assistance via short pulses \cite{SchutzholdGiesDunne2008},
use precisely the same IHO mathematics.

\subsection{Horizons: Rindler, Schwarzschild, de Sitter}

Near the Schwarzschild horizon in Kruskal coordinates, the outgoing modes scale as $U \propto e^{-\kappa t}$ and the ingoing modes as $V \propto e^{\kappa t}$, where $\kappa = 1/4GM$ is the surface gravity. These are the IHO trajectories $Q(t)=Q_{0}e^{\omega t}$, $P(t)=P_{0}e^{-\omega t}$ under the identification $\omega \to \kappa$, expressed in the natural coordinate system at the horizon. The near-horizon boost generator is the Hamiltonian conjugate to the Killing time of an asymptotic observer that has the Berry--Keating form \eqref{eq:HBK}. The Hawking temperature \cite{Hawking1975,KumarMartoPTEP2024} emerges from the IHO thermal Bogoliubov transformation; the Unruh effect for an accelerating observer \cite{Unruh1976} is the same calculation in Rindler space. In black-hole perturbation theory, the peak of the effective potential barrier (near $r=3M$ for the Schwarzschild geometry) is locally a saddle, and the discrete tower of complex quasinormal-mode frequencies that governs the ringdown of a perturbed black hole \cite{Chandrasekhar1983, Berti2009} is the spectrum of the IHO under outgoing-only boundary conditions.

The same structure operates at the cosmological horizon of de Sitter spacetime. The static-patch observer sees a thermal spectrum at the Gibbons--Hawking temperature \cite{GibbonsHawking1977} for exactly the same IHO-Bogoliubov reason that an inertial Minkowski observer sees a Planck spectrum at the surface of a black hole.

\subsection{Inflation: the cosmological IHO}

The Mukhanov--Sasaki mode equation \cite{Mukhanov1992,GaztanagaKumarMartoCQG2026} for inflationary curvature perturbations, after canonical rescaling, takes the form of an inverted oscillator in conformal time,
\begin{equation}
\label{eq:MS}
\Phi_{k}^{\prime\prime} + \Bigl[k^{2} - \tilde\mu_{\mathrm{eff}}^{2}(\tau)\Bigr]\Phi_{k} \;=\; 0, \qquad \tilde\mu_{\mathrm{eff}}^{2}(\tau) \;=\; \frac{2}{\tau^{2}} \;>\; 0,
\end{equation}
where the effective mass squared is \emph{negative} (the term enters \eqref{eq:MS} with a minus sign). Modes crossing the Hubble horizon are not the quiet freeze-out of a stable oscillator; they are the exponential amplification of an inverted one. Every primordial fluctuation that has produced an observable CMB temperature anisotropy passed through this IHO stage during the few $e$-folds when its Fourier wavelength stretched across the Hubble scale. The amplitude of the primordial power spectrum, its spectral tilt, and the tensor-to-scalar ratio are all set by IHO dynamics during horizon exit. The CMB is, in this sense, the largest-screen movie ever made of the inverted oscillator at work.

\subsection{Mathematics: Berry, Keating, and the Riemann zeros}

The most surprising appearance of the IHO is in pure mathematics. In 1999, Michael Berry and Jonathan Keating observed \cite{BerryKeating1999, BerryKeating1999SIAM} that if a self-adjoint Hamiltonian existed whose spectrum equals the imaginary parts of the non-trivial zeros of the Riemann zeta function, then the Riemann hypothesis would follow as a corollary. Their candidate was the dilatation Hamiltonian $H_{BK} = (QP+PQ)/2$ which is the IHO in Berry--Keating form. Under a particular phase-space quantization condition with hard cutoffs $|Q|\geq\ell_{Q}$, $|P|\geq\ell_{P}$ obeying $\ell_{Q}\ell_{P}=2\pi\hbar$, the state count produced by $H_{BK}$ reproduces the Riemann--von Mangoldt asymptotic density of zeros,
\begin{equation}
N(E) \;\simeq\; \frac{E}{2\pi}\Bigl[\ln\!\frac{E}{2\pi}\,-\,1\Bigr] \;+\; \frac{7}{8},
\end{equation}
exactly. Alain Connes \cite{Connes1999} arrived at a closely related structure through noncommutative-geometric methods. Germ\'an Sierra \cite{Sierra2008} gave an explicit field-theoretic realization through a cyclic renormalization-group flow. The connection between the IHO and the Riemann zeros at the level of the spectral density is, in fact, older still: Bhaduri, Khare, and Law \cite{BhaduriKhareLaw1995} showed in 1995 that the smooth phase of $\zeta(\tfrac{1}{2}+iE)$ along the critical line is the density of states of an inverted oscillator. The full program has since been reviewed at colloquium level \cite{SchumayerHutchinson2011}.

The long-standing puzzle in this program has been the absence of a geometric reason for the boundary condition that selects the Riemann zeros. In the direct-sum framework developed jointly with Gazta\~naga and Marto in our work on Einstein–Rosen bridges \cite{GaztanagaKumarMartoCQG2026}, this boundary condition emerges naturally as a matching across $\mathcal{PT}$-conjugate phase-space sectors. We will return to this in Section~\ref{sec:KL}. The point to emphasize here is not the prospect of resolving the Riemann hypothesis through physical reasoning which still remains an open and difficult problem, but the more modest and striking fact that the same dilatation Hamiltonian that governs horizons, LLL scattering, and the additional spin–2 sector of quantum (quadratic) gravity also appears in the deepest known spectral problem of number theory. This is not the kind of coincidence physics produces by accident.

A recent thread connects the IHO to the supercritical inverse-square potential through
an exact spectral duality \cite{SundaramBurgessODell2024}: both share \eqref{eq:HBK}
as an intermediate stage, and the boundary-condition flow of the inverse-square
potential is a quantum renormalization-group limit cycle, one of the cleanest examples
of cyclic RG anomaly. The IHO inherits this cyclic structure as a feature of the family
of self-adjoint extensions one can choose for it. The inverse-square potential is itself
a recurring object across physics, describing the electron--polar-molecule interaction
\cite{FermiTeller1947}, the Efimov effect in three-body bound states predicted in nuclear
physics \cite{Efimov1970} and observed in ultracold atoms \cite{Kraemer2006}, and the
exactly solvable Calogero--Sutherland many-body problem with pairwise inverse-square
couplings \cite{Calogero1971, Sutherland1971}.

\subsection{Why the IHO has been hidden in plain sight}

If the IHO appears so widely, why is it not a standard tool of the discipline alongside the harmonic oscillator? Three reasons converge.

First, the IHO has no normalizable vacuum, and twentieth-century quantum field theory is built on Fock spaces above normalizable vacua. Systems without such vacua have been routinely relegated to the periphery of ``not really QFT,'' as the implicit attitude has it. This is more a habit of the discipline than a physical principle.

Second, the IHO is dynamically unstable. Wavepackets spread; correlations grow exponentially; standard $S$-matrix theory, which assumes asymptotic free states, struggles to characterize it natively. But every horizon, every transition state, every critical point is dynamically unstable; the difficulty is with the orthodox formalism, not with the physics. Quantum field theory in curved spacetime has been quietly handling such systems for fifty years, and the foundational problem of the absence of a privileged vacuum rather than being an exotic complication is the generic situation outside Minkowski space.

Third, the IHO field has been culturally associated with pathology. The ghost interpretation of higher-derivative gravity, the bad reputation of tachyons, the warnings about negative-frequency modes, all of these have clustered into an inherited sense that any time an ``inverted'' kinetic or mass term appears, theoretical trouble is at hand. The IHO is structurally distinct from the ghost, from the tachyon, and from a propagating negative-frequency mode. It is its own object, with its own physics, and that physics has been hiding in the foundations.

The remainder of this paper applies the IHO to the one place in physics where its presence has been most consequentially missed: the spin--2 sector of the unique renormalizable theory of gravity.

% ====================================================
\section{Renormalizability as a physical principle}
\label{sec:renormalizability}

Before the renormalization revolution, ultraviolet divergences looked like evidence that quantum field theory was internally diseased. Landau's zero-charge problem in QED \cite{LandauPomeranchuk1955} suggested that the theory could not exist at arbitrarily short distances. Dirac \cite{DiracDirections1978} objected to the conceptual meaning of renormalization. Heisenberg \cite{Heisenberg1943} proposed an $S$-matrix theory in which local fields were abandoned in favour of observable scattering amplitudes. Chew's bootstrap \cite{Chew1962} pushed the attitude further: hadrons would be determined by self-consistency of the $S$-matrix rather than by more fundamental fields, and the Veneziano amplitude \cite{Veneziano1968} showed how rich that road could be. In parallel, current algebra and Partial Conservation of the Axial Current (PCAC) \cite{GellMannLevy1960, AdlerDashen1968} extracted low-energy theorems from symmetry without specifying the ultraviolet completion. Fermi's four-fermion theory \cite{Fermi1934} described $\beta$-decay with striking success, but its coupling carried mass dimension $-2$ and so announced its own breakdown at $\sqrt{s}\sim 300\,\mathrm{GeV}$.

What changed between 1948 and 1973 was not that divergences disappeared. It was that physicists learned to classify them. A four-dimensional interaction with coupling $g$ is perturbatively renormalizable when $[g]\geq 0$. Loop integrals may diverge, but the divergent structures are already present in the original Lagrangian, and a finite number of measurements fixes a finite number of parameters. The theory then makes predictions. Dyson \cite{Dyson1949} showed that all divergences of QED at all loop orders reduce to three primitives. 't~Hooft \cite{tHooft1971} proved that spontaneously broken non-Abelian gauge theories are renormalizable to all orders. Wilson \cite{Wilson1971} reframed renormalizability itself as a Wilsonian flow on coupling space, turning a calculational nuisance into a physical principle about which short-distance degrees of freedom survive coarse-graining. Gross, Politzer, and Wilczek \cite{GrossWilczek1973, Politzer1973} discovered asymptotic freedom and ended the bootstrap program within three years. Renormalizability was not a mathematical convenience. It was a falsification criterion which is the sharpest predictive instrument theoretical physics had produced and it told physicists when an interaction could be fundamental and when it was an effective approximation waiting for its ultraviolet completion.

General relativity, treated perturbatively from the Einstein--Hilbert action alone, fails this criterion. The gravitational coupling $\kappa \sim \MP^{-1}$ has negative mass dimension, and new counterterms appear at successively higher loop orders. As an effective field theory below the Planck scale \cite{Donoghue1994}, this is not a contradiction. As a fundamental theory, it is a warning: the ultraviolet completion has not yet been included. The non-renormalizability of perturbative GR is, in this respect, structurally identical to the non-renormalizability of Fermi's four-fermion theory in 1933. It is a quantitative arrow pointing at new physics. The historical question is what new physics it points at.

% ====================================================
\section{Stelle's theorem and the spin--2 fork}
\label{sec:stelle}

In 1977, K.\ S.\ Stelle proved that adding the two independent curvature-squared invariants to the Einstein--Hilbert action makes gravity perturbatively renormalizable in four dimensions \cite{Stelle1977, Stelle1978}. The action,
\begin{equation}
\label{eq:Stelle}
S \;=\; \int d^{4}x\,\sqmg\,\left[\,\frac{\MP^{2}}{2}\,R \;+\; \frac{\alpha}{2}\,R^{2} \;+\; \frac{\beta}{2}\,\Wmnrs W^{\mu\nu\rho\sigma}\,\right],
\end{equation}
is, up to the Gauss--Bonnet identity, field-redefinition choices, and the inclusion of a cosmological constant, the local curvature-squared completion of general relativity in four dimensions. The $R^{2}$ term introduces a propagating scalar called "the scalaron" whose dynamics in the Einstein frame produce the Starobinsky inflationary potential \cite{Starobinsky1980}. The Weyl-squared term improves the ultraviolet falloff of the spin--2 propagator from $1/k^{2}$ to $1/k^{4}$. Loop divergences close on the same finite set of couplings, and the theory was subsequently shown to be asymptotically free in the gravitational sector by Fradkin and Tseytlin \cite{FradkinTseytlin1982} and by Avramidi and Barvinsky \cite{AvramidiBarvinsky1985} which provide a stronger ultraviolet status than QED itself enjoys. By the same criterion that singled out the renormalizable structure of QED, the electroweak theory, and QCD, gravity had acquired a candidate ultraviolet completion. By the standards that ended the bootstrap program in three years, this should have been a turning point.

It was not. The reason was the second spin--2 pole. With signature $\eta_{\mu\nu}=(-,+,+,+)$, in which timelike four-momentum satisfies $k^{2}<0$ and spacelike four-momentum satisfies $k^{2}>0$, the spin--2 part of the propagator around Minkowski takes the schematic form
\begin{equation}
\label{eq:propspin2}
D^{(2)}_{\mu\nu,\rho\sigma}(k) \;=\; -\,\frac{i}{\MP^{2}}\,P^{(2)}_{\mu\nu,\rho\sigma}\left[\,\frac{1}{k^{2}} \;-\; \frac{1}{k^{2}-\MP^{2}/\beta}\,\right].
\end{equation}
The location of the second pole depends on the sign of $\beta$:
\begin{equation}
\label{eq:signfork}
\frac{\MP^{2}}{\beta} \;=\;
\begin{cases}
-\,m_{2}^{2},\quad \beta<0,\;m_{2}^{2}>0,\\[3pt]
+\,\mu_{2}^{2},\quad \beta>0,\;\mu_{2}^{2}>0.
\end{cases}
\end{equation}

For $\beta<0$, the pole is timelike which represents a real massive spin--2 particle that appears with the wrong residue. Quantized in the standard Fock space representation, this is the well-known massive spin–2 ghost. For half a century, the dominant attempts to live with this fact have tried to retain the $\beta<0$ branch while preventing the ghost from appearing as an external state: Lee--Wick complex-contour quantization \cite{LeeWick1969}, the fakeon prescription (in Euclidean signature) in which the ghost is removed from the asymptotic Hilbert space by fiat at the cost of microcausality \cite{Anselmi2017, AnselmiPiva2018}; PT-symmetric and Mannheim-type modifications of the inner product \cite{Bender2007, Mannheim2018}; non-perturbative QCD-analogue reshuffling \cite{HoldomRen2016}; the scale-invariant ``agravity'' completion of Salvio and Strumia \cite{SalvioStrumia2014, Salvio2018}; the super-renormalizable and nonlocal infinite-derivative extensions that replace polynomial higher derivatives with entire functions \cite{Modesto2012, BuoninfanteLambiaseMazumdar2019}. Each program is technically substantial, and each has produced genuine progress. What they share is a starting premise: that the additional sector is a ghost, and that the task of a consistent quantum gravity is to suppress it. We will argue here that the premise itself is the obstruction.

For $\beta>0$, the pole is at spacelike momentum. The second spin--2 sector is not an ordinary massive particle. After diagonalisation of the transverse-traceless sector following \cite{KumarMartoQQG2026, KumarMartoUnitarity2026}, one obtains
\begin{equation}
\label{eq:HTTdiag}
H_{TT} \;=\; \tfrac{1}{2}\sum_{s}\!\int\!\tfrac{d^{3}k}{(2\pi)^{3}}\bigl[\dot u_{s}^{\,2}+k^{2}u_{s}^{\,2}\bigr] \;-\; \tfrac{1}{2}\sum_{s}\!\int\!\tfrac{d^{3}k}{(2\pi)^{3}}\bigl[\dot v_{s}^{\,2}+(k^{2}-\mu_{2}^{2})v_{s}^{\,2}\bigr],
\end{equation}
where $u_{s}$ is the massless graviton and $v_{s}$ is the additional spin--2 field. Expanding $v_{s}$ and its conjugate momenta 
in eigenmodes of $-\nabla^{2}f_{\lambda}=\lambda f_{\lambda}$
\begin{equation}
v_s(t,\vec x)
=
\sum_\lambda q_{s,\lambda}(t) f_\lambda(\vec x),
\qquad
\pi_s(t,\vec x)
=
\sum_\lambda p_{s,\lambda}(t) f_\lambda(\vec x).
\end{equation}
gives the mode Hamiltonian
\begin{equation}
\label{eq:Hslambda}
H_{s,\lambda} \;=\; -\,\tfrac{1}{2}p_{s,\lambda}^{\,2} \;+\; \tfrac{1}{2}(\mu_{2}^{2}-\lambda)\,q_{s,\lambda}^{\,2},
\end{equation}
which is not a negative harmonic oscillator. It is an indefinite hyperbolic system---a dual IHO. The kinetic term carries the opposite sign from a conventional oscillator and the ``mass'' term is positive. The classical motion is hyperbolic (right panel of Figure~\ref{fig:iho-dual}), the spectrum is real and continuous, and the canonical quantization has no normalizable vacuum.

The crucial conceptual point, developed in detail in Section~\ref{sec:opening}, is that this is not a ghost. A ghost is a negative-norm Hilbert-space excitation; the dual IHO has no such particle excitation to assign a negative norm to. The standard slogan of ``wrong-sign kinetic term means ghost'' is wrong as a diagnosis. It conflates two distinct mathematical situations that share a notation but not a physics.

% ====================================================
\section{The K\"all\'en--Lehmann theorem and unitarity}
\label{sec:KL}

If the spin--2 sector is a dual IHO, the foundational question is how a quantum field theory can contain a sector that has no Fock vacuum. The answer is not to pretend the IHO is a particle. The answer is to formulate the theory in the Hilbert-space structure the IHO naturally demands \cite{GaztanagaKumarMartoCQG2026}.

\begin{figure}
    \centering
    \includegraphics[width=0.5\linewidth]{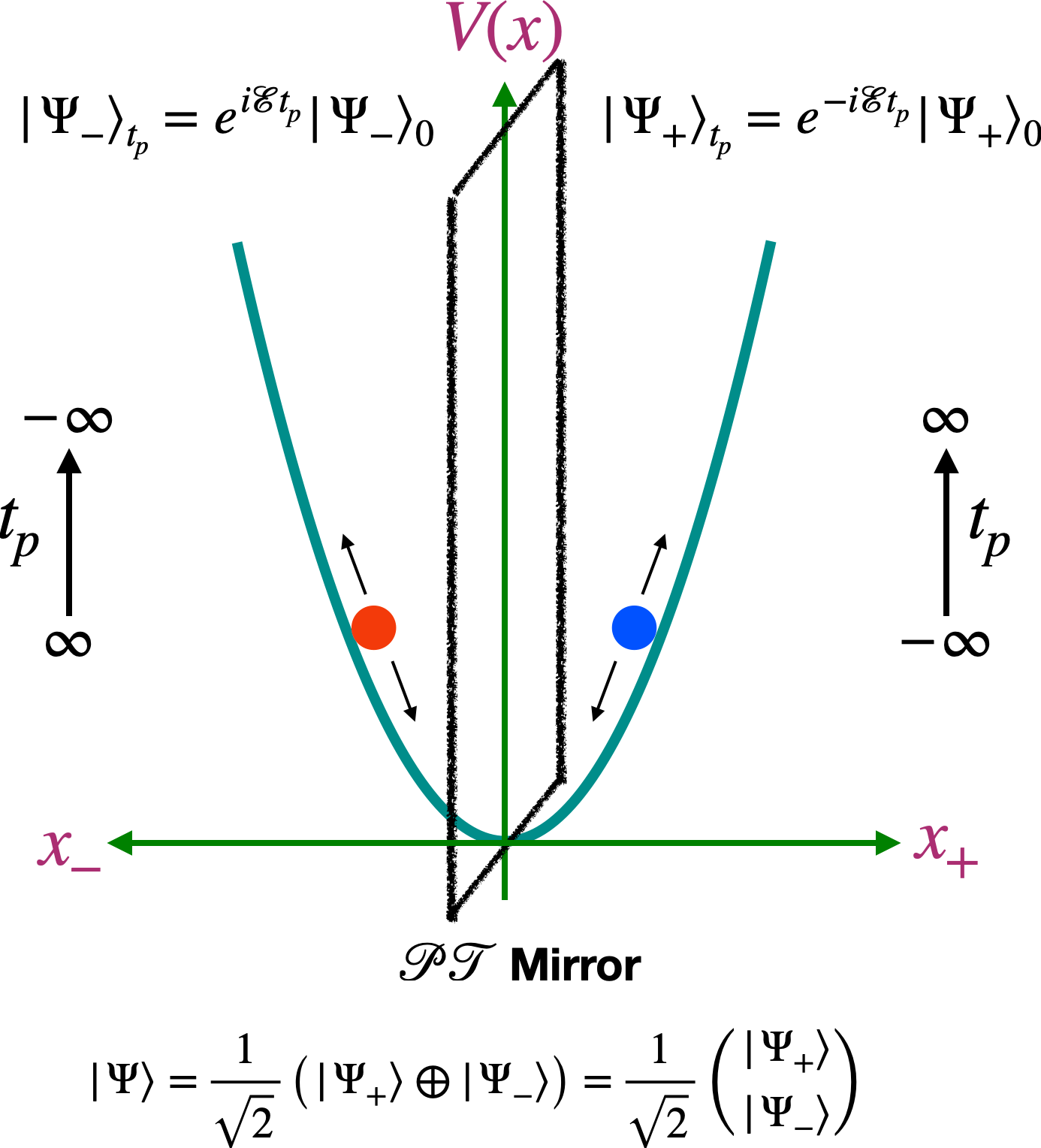}
    \caption{Direct-sum quantum mechanics on the harmonic potential $V(x)$, showing how a single classical system carries two $\mathcal{PT}$-conjugate quantum sectors. The vertical axis at $x=0$ acts as a $\mathcal{PT}$ mirror, partitioning configuration space into right ($x_+$) and left ($x_-$) sectors. Each sector hosts an independent wavefunction $|\Psi_\pm\rangle$ evolving with its own time parameter $t_p$ along opposite arrows: $|\Psi_+\rangle_{t_p} = e^{-i\mathcal{E}t_p}|\Psi_+\rangle_0$ and $|\Psi_-\rangle_{t_p} = e^{+i\mathcal{E}t_p}|\Psi_-\rangle_0$. The full physical state is the direct sum $|\Psi\rangle = (|\Psi_+\rangle \oplus |\Psi_-\rangle)/\sqrt{2}$, with sectors superselected and no cross-sector matrix elements for local observables. For the harmonic oscillator this construction is harmless, since the compact orbits and normalizable vacuum make a single-sector quantization equally consistent. For the IHO it is forced: the hyperbolic flow itself distinguishes the two arrows of time, and the four-region phase-space structure of Figure~\ref{fig:iho-dual} is the geometric origin of this direct-sum.}
    \label{fig:HODQFT}
\end{figure}

The (dual-)IHO phase space is divided by separatrices into four regions related by parity-time reflections, and opposite regions carry opposite arrows of time (Figure~\ref{fig:iho-dual}). A single Schr\"odinger equation imposes one arrow by hand. This is harmless for the harmonic oscillator, whose orbits are compact and whose vacuum is normalizable; it is not harmless for the IHO, where the hyperbolic flow itself forces the distinction between conjugate time orientations. Direct-sum quantum field theory (DQFT) \cite{KumarMartoUniverse2024, KumarMartoPTEP2024, KumarMartoGRG2024, KumarMartoSymmetry2025, GaztanagaKumarJCAP2024, GaztanagaKumarMartoCQG2026, GaztanagaKumarSymmetry2025} was developed to handle this situation. A quantum state is written as
\begin{equation}
\label{eq:directsum}
|\Psi\rangle \;=\; \tfrac{1}{\sqrt{2}}\bigl(|\Psi_{+}\rangle \,\oplus\, |\Psi_{-}\rangle\bigr),\qquad \mathcal{H} \;=\; \mathcal{H}_{+}\oplus\mathcal{H}_{-},
\end{equation}
with $\mathcal{PT}$-conjugate sectors carrying opposite arrows of time and no cross-sector mixing (See Fig.~\ref{fig:HODQFT} for the schematic understanding on quantum harmonic oscillator in this formulation)

The direct-sum Schr\"odinger equation has the structure
\begin{equation}
i\partial_{t}\!\begin{pmatrix}|\Psi_{+}\rangle\\ |\Psi_{-}\rangle\end{pmatrix}
\;=\;
\begin{pmatrix}\hat H_{+} & 0\\ 0 & -\hat H_{-}\end{pmatrix}\!
\begin{pmatrix}|\Psi_{+}\rangle\\ |\Psi_{-}\rangle\end{pmatrix},
\end{equation}
The Hilbert space splits accordingly into a geometric superselection of sectors associated with regions of spacetime related by discrete transformations. 
Similarly all field operators in DQFT are split into direct-sum of two components 
\begin{equation}
    \hat \phi =  \frac{1}{\sqrt{2}}\left( \hat\phi_+\oplus \phi_- \right)= \frac{1}{\sqrt{2}}\begin{pmatrix}
        \hat \phi_+ & 0 \\ 
        0 & \hat\phi_- 
    \end{pmatrix}
\end{equation}
corresponding to the direct-sum Fock space. 
The field operators corresponding to opposite arrows of time commute. For example, for a scalar field one has, in particular, $[\hat\phi_{+}(x),\hat\phi_{-}(-y)]=0$ in addition to the usual spacelike commutators within each sector.  This is the Hilbert-space environment in which the IHO finds its natural home: its four phase-space regions become superselection sectors of a direct-sum state space, and the missing ``geometric reason'' for the Berry--Keating boundary condition, which has been identified as an open problem in mathematical physics since 1999, becomes the matching condition across these sectors.

The central technical claim of the present program can now be stated as a theorem. It is short, and it is the engine that drives the rest of this paper.

\begin{theorem}[Dual-IHO spectral theorem]
\label{thm:KL}
In the $\beta>0$ branch of local quadratic gravity, the additional spin--2 sector has vanishing physical K\"all\'en--Lehmann spectral density. Its propagator therefore contains no absorptive $\delta(k^{2}-\mu_{2}^{2})$ contribution and is uniquely fixed, in the physical spectral representation, to its Cauchy principal-value form.
\end{theorem}

\noindent\emph{Argument.} For a standard field with a normalizable vacuum and physical one-particle spectrum, the K\"all\'en--Lehmann representation \cite{Kallen1952, Lehmann1954} reads
\begin{equation}
\label{eq:KL}
\widetilde\Delta(k^{2}) \;=\; \frac{-iZ}{k^{2}+m^{2}-i\epsilon} \;+\; \int_{0}^{\infty}\!ds\,\frac{-i\,\rho(s)}{k^{2}+s-i\epsilon}, \qquad
\rho(s) \;=\; \sum_{n}\bigl|\langle 0|\phi(0)|n\rangle\bigr|^{2}\,\delta(s-M_{n}^{2})\,\geq\,0.
\end{equation}
Here \(Z\) is the field-strength renormalization constant, or equivalently the residue of the physical one-particle pole. It measures how strongly the field operator \(\phi\) creates the one-particle state from the vacuum:
\[
Z = \left|\langle 0|\phi(0)|1\rangle\right|^2 .
\]
Thus \(Z>0\) means that the propagator contains a normal physical particle pole, while \(Z=0\) means that no such physical one-particle state contributes.
The lower limit $s\geq 0$ is not a convention but the Wightman spectrum condition: physical states carry timelike or null four-momentum with non-negative invariant mass squared. The dual-IHO spin--2 sector fails to contribute to this physical spectral measure for two independent reasons.

First, the dual IHO has no normalizable $L^{2}$ vacuum, so the matrix element $\langle 0|\phi(0)|n\rangle$ defining a positive spectral weight over physical particle states is absent. The eigenfunctions of \eqref{eq:Hslambda} are parabolic-cylinder functions that grow exponentially in one direction, and the spectral sum in \eqref{eq:KL} has no positive contribution to assemble.

Second, the pole of the dual-IHO propagator sits at $k^{2}=+\mu_{2}^{2}>0$, which would correspond in \eqref{eq:KL} to $s=-\mu_{2}^{2}<0$, outside the physical spectral domain. No physical particle with real mass can produce a spacelike pole. Each of these reasons, individually, would force the spectral weight to vanish at the would-be pole. Together they overdetermine the conclusion:
\begin{equation}
\label{eq:rhozero}
\rho_{\dIHO}(s) \;=\; 0.
\end{equation}
With $\rho=0$ in the spectral representation, the would-be $\delta(k^{2}-\mu_{2}^{2})$ piece of the propagator is identically absent. The propagator retains only its dispersive part,
\begin{equation}
\label{eq:PVprop}
\Delta_{\dIHO}(k) \;=\; \PV\!\left[\frac{i}{k^{2}-\mu_{2}^{2}}\right].
\end{equation}
This principal value is not imposed by analogy with fakeons \cite{Anselmi2017}, Lee--Wick contours \cite{LeeWick1969}, or a Feynman--Wheeler average \cite{AnselmiPiva2018}. It is forced by the absence of physical spectral support.

The optical theorem follows immediately. The absorptive part of any forward amplitude is determined by Cutkosky cuts \cite{Cutkosky1960} that put internal lines on shell. A massive timelike ghost would contribute to the cut sum with the wrong sign and so violate the unitarity bound. The dual IHO contributes nothing at all: kinematic cuts have causal support $k^{0}>0$, $k^{2}\leq 0$, whereas the dual-IHO on-shell condition $k^{2}=\mu_{2}^{2}>0$ is spacelike. Hence
\begin{equation}
\label{eq:Cutkosky}
\mathrm{Cut}_{\dIHO} \;=\; 0,\qquad
2\,\mathrm{Im}\,\mathcal{M}(a\to a) \;=\; \sum_{f\in\{\mathrm{graviton},\,\mathrm{scalaron}\}}\!\!\int\!d\Pi_{f}\,|\mathcal{M}(a\to f)|^{2}.
\end{equation}
The physical absorptive part is saturated by the massless graviton and the scalaron. The dual-IHO spin--2 mode participates only in off-shell virtual exchange and contributes to ultraviolet renormalization. The ultraviolet improvement of the Stelle propagator, $D^{(2)}(k)\sim \mu_{2}^{2}/k^{4}$ at large $k$, is preserved by the principal-value structure, because the difference between an ordinary Feynman propagator and a principal-value distribution is exactly the $\delta$-function piece forbidden by \eqref{eq:rhozero}. Renormalizability is intact; unitarity is intact; the central technical objection that drove the field away from quadratic gravity for half a century has been answered.

The intuitive content of Theorem~\ref{thm:KL} is simple. A negative-residue pole at timelike momentum is a ghost. The same pole at spacelike momentum, in a sector with no normalizable vacuum, is not a ghost; it is a dual IHO. 

% ====================================================
\section{Cosmology, finite action, and falsifiability}
\label{sec:cosmology}

A theory of quantum gravity must speak to the Universe, not only to its own internal consistency. The reformulation developed here makes contact with cosmology in three connected ways \cite{KumarMartoQQG2026}: inflation, parity-asymmetric large-scale correlations \cite{GaztanagaKumarJCAP2024,GaztanagaKumarSymmetry2025}, and singularity avoidance.

The scalaron of the $R^{2}$ sector gives the Starobinsky inflationary potential, currently sitting at the top of every CMB likelihood \cite{Planck2018Inflation}. In the unitary $\beta>0$ branch the dual-IHO spin--2 sector does not appear as an extra particle in the CMB; it remains virtual. It does, however, renormalize the tensor sector through off-shell virtual exchange, and the resulting tensor-to-scalar ratio is enhanced over the Starobinsky baseline by a calculable factor:
\begin{equation}
\label{eq:rformula}
r \;\simeq\; \frac{12}{N^{2}}\cdot\frac{12\alpha}{12\alpha-\beta}, \qquad 12\alpha>\beta>0.
\end{equation}
The Starobinsky value $r_{0}\simeq 12/N^{2}\approx 4\times 10^{-3}$ (for $N\approx 55$) is recovered as $\beta\to 0$. For natural values of $\beta$ at the scale of inflation, $\beta\sim 10^{9}$, the enhancement reaches the percent level which is a small effect, but one within the projected sensitivity of LiteBIRD \cite{LiteBIRD2023} and CMB-S4 \cite{CMBS42019}. These experiments are, in effect, a test of the Weyl coefficient of the unique renormalizable theory of gravity.

The direct-sum quantization of inflationary fluctuations produces a parity-asymmetric primordial power spectrum. The two $\mathcal{PT}$-conjugate sectors pick up opposite-sign slow-roll corrections, $\nu_{s}^{\pm}\simeq \tfrac{3}{2}\pm 2\epsilon_{1}$, and the resulting angular power spectrum carries an alternation between odd and even multipoles,
\begin{equation}
\label{eq:Cparity}
C_{\ell}^{\mathrm{DSI}} \;=\; C_{\ell}^{\mathrm{SI}}\,\bigl[\,1+(-1)^{\ell+1}\,\Delta C_{\ell}\,\bigr],
\end{equation}
with odd $\ell$ enhanced and even $\ell$ suppressed at the level of a few percent (See \cite{GaztanagaKumarJCAP2024,GaztanagaKumarSymmetry2025,GaztanagaKumarMartoCQG2026} for more details). This matches, without further adjustment, the well-known low-$\ell$ parity anomaly of the CMB \cite{Land2005, KimNaselsky2010,GaztanagaKumarJCAP2024,GaztanagaKumarSymmetry2025}: the standard parity indicator $R_{TT}(\ell_{\max})$ sits at approximately $0.79$ in the Planck data for $\ell\leq 20$--$30$, and the direct-sum prediction recovers this value. The likelihood analysis against Planck 2018 \cite{GaztanagaKumarMartoCQG2026, GaztanagaKumarJCAP2024, GaztanagaKumarSymmetry2025} gives a Bayesian preference of approximately $650$:$1$ in favor of the direct-sum spectrum over the standard inflationary one. Observing similar parity asymmetry in the $B-$mode polarization sector could further validate the framework of direct-sum inflation \cite{GaztanagaKumarMartoCQG2026,KumarMartoQQG2026}.

The same framework supports the finite-action principle of Barrow, Lehners, and Stelle \cite{BarrowLehners2020, LehnersStelle2019}, which demands that physically admissible cosmological histories have finite on-shell Euclidean action as $t\to 0$. For the Weyl-squared term this is a strong constraint: anisotropic Bianchi-IX (BKL/Mixmaster) singularities \cite{BKL1970}, in which the Weyl scalar diverges as $1/t^{4}$ while the spatial volume vanishes only linearly, produce a polynomially divergent action,
\begin{equation}
W^{2}\sim \tfrac{1}{t^{4}},\quad \sqmg\sim t \;\Longrightarrow\; S_{W} \sim \int dt\,\sqmg\,W^{2} \sim \int dt\,\tfrac{1}{t^{3}},
\end{equation}
and are excluded from the gravitational path integral. The path integral concentrates instead on histories that approach an isotropic FLRW patch near the initial moment, with the dual IHO acting as the dynamical channel that bleeds anisotropy out of the geometry. From this, Starobinsky inflation emerges not as a model imposed by hand but as the preferred late-time state of a quantum-cosmological initial condition. The Schwarzschild interior is related to an anisotropic Kantowski--Sachs cosmology which suggests an analogous exclusion of black-hole singularities. Whether the program can be completed to a fully singularity-free quantum gravity is an open question; the calculations are tractable and the mathematical infrastructure is in place.

% ====================================================
\section{The historical position of contemporary alternatives}
\label{sec:alternatives}

If the argument so far is right, the larger conclusion is implicit. Each of the major contemporary programs in quantum gravity has produced genuine mathematical and physical insight. The AdS/CFT correspondence \cite{Maldacena1998} is among the most powerful computational tools ever discovered in theoretical physics, and the holographic entropy bounds it has yielded for black-hole microstate counting are independent of any commitment to string theory as a fundamental description. The asymptotic-safety program, originally proposed by Weinberg \cite{Weinberg1979}, has produced non-trivial evidence in carefully truncated theory spaces for a UV fixed point of the gravitational renormalization group \cite{ReuterSaueressig2019}. Loop quantum gravity has clarified the kinematic structure of background-independent quantization in ways that have informed condensed-matter analogues and tensor-network constructions \cite{Rovelli2004}. The on-shell amplitudes program has reorganized our understanding of perturbative scattering \cite{Cheung2018}. These contributions are real.

What none of these programs has produced is what the renormalization criterion asks for: a Lagrangian with finitely many parameters, in Lorentzian signature, with a unitary Hilbert space, whose loop integrals close on themselves. Each has taken the structural position of one of the speculative frameworks that the renormalization revolution superseded between 1948 and 1973. String theory, with its insistence on amplitudes rather than a fundamental local Lagrangian and its landscape of vacua \cite{Susskind2003}, occupies the position the bootstrap occupied in the 1960s. Asymptotic safety, with its replacement of a Lagrangian by a self-consistency condition at a fixed point, occupies the position of dispersion relations. Loop quantum gravity occupies the position of current algebra without QCD. The amplitudes program occupies the position of Regge theory before QCD. The ``GR as an effective field theory'' position \cite{Donoghue1994}, defended for thirty years, occupies the position of ``Fermi theory as an EFT'' before the $W$ boson, except that, in the gravitational case, the UV completion of the EFT is supposed to point at \emph{already exists, and has existed for forty-eight years}.

That last clause is the difference between then and now. In the 1930s and 1940s the alternatives to renormalized field theory were honest: the electroweak theory had not yet been discovered, the bootstrap was a serious hypothesis because there was no other game in town, and Heisenberg, Chew, Landau were responding to a problem whose solution was genuinely unknown. The discovery of QCD and electroweak theory ended that era cleanly: the alternatives were not refuted by argument but rendered obsolete by a better theory. Today, the better theory exists. Stelle's action is not a conjecture; it is a Lagrangian, renormalizable to all orders, that contains general relativity as its long-wavelength limit, the Starobinsky inflationary potential as its scalar sector, a unitary quantization once the dual-IHO reformulation is admitted, and concrete CMB predictions. The community has not adopted it because it learned to live without it during decades when the ghost objection was assumed fatal. The objection is no longer fatal. The reason for the deferral is no longer the reason.

% ====================================================
\section{Conclusion: the saddle point as the road back}
\label{sec:conclusion}

The harmonic oscillator built modern physics. The Fock-space formalism of perturbative quantum field theory, the spectroscopy of atoms, rotational energies of molecules, black body radiation, the structure of the QED and QCD vacuum at high energy, the entire predictive apparatus of the Standard Model, all of these ride on the universality of the harmonic oscillator as the leading-order description of stable equilibrium. It is one of the most successful elementary structures in the history of science.

The inverted harmonic oscillator is the quiet counterpart. It is what physics writes down at the saddle point of every barrier, the symmetric point of every Mexican hat, the dividing surface of every chemical reaction, the lowest Landau level of quantum-Hall point contact, the horizons of black hole and de Sitter universe, Rindler horizons, Schwinger pair production, the freeze-out of every inflationary mode, and the Berry–Keating Hamiltonian's route to the Riemann hypothesis. It has been there all along, doing universal work in physics, while being categorized as exotic by the formal apparatus of QFT that prefers normalizable vacua and stable particles. The argument of this paper is that the same elementary structure also sits, structurally, at the heart of the unique perturbatively renormalizable theory of gravity in four dimensions. The ghost diagnosis that drove the community away from Stelle's action in 1977 was a misidentification: the additional spin--2 sector, with the natural sign $\beta>0$, is not a ghost but a dual IHO spin--2 field, with no normalizable vacuum, spacelike spectral support, and vanishing physical K\"all\'en--Lehmann spectral density. Its propagator is principal-value, not because we choose it to be, but because there is no absorptive spectral weight to add. Unitarity and renormalizability coexist in the original local theory.

This is not a retreat from quantum field theory. It is a correction to the assumption that every quantum field must be understood as a tower of stable oscillator particles above a single normalizable vacuum. The harmonic oscillator built perturbative QFT. The inverted harmonic oscillator is what QFT meets at horizons, at saddle points, and in the early Universe. A theory of quantum gravity that contains both is not less quantum. It is more faithful to the systems gravity actually gives us.

The deepest reason to give this construction the attention it asks for is the one this paper has tried to make audible. The IHO is the saddle point of physical change that we can witness across multiple physical domains. To find the same structure at the center of the unique renormalizable theory of gravity is to find that gravity speaks the same elementary language that the rest of physics already speaks. The community does not need another beautiful escape from predictivity. It needs to return to the criterion that built quantum field theory and read its gravitational answer without the inherited prejudice.
The answer has been waiting since 1977. The ghost was the wrong consideration. The saddle point has been the right one all along.

\bigskip

\paragraph{Acknowledgements.} The author thanks J.\ Marto, E.\ Gazta\~naga, David Wands, L.\ Buoninfante, and colleagues at the Institute of Cosmology and Gravitation, University of Portsmouth, for discussions that shaped the perspective presented here. Support from a Royal Society Newton International Fellowship (2023-25) is gratefully acknowledged.

% ============================================================
%  Bibliography
% ============================================================
\bibliographystyle{utphys}
\bibliography{refs}

\end{document}